\documentclass[preprint,preprintnumbers,nofootinbib,aps,10pt,twocolumn]{revtex4-1}
\usepackage{amsmath,amssymb,bm,epsfig}
\usepackage{color}
\usepackage{natbib}
\usepackage{hyperref} 
\usepackage{ulem}
\usepackage{graphicx}
\usepackage{xcolor,colortbl}
\RequirePackage{lineno}
\newcommand{\sNN}{\sqrt{s_{\textrm{NN}}}}

\newcommand{\prp}{\textrm{p}}
\newcommand{\pbar}{\overline{\rm p}}

\definecolor{Gray}{gray}{0.85}
\newcolumntype{a}{>{\columncolor{Gray}}c}

\def \beq{\begin{equation}}
\def \eeq{\end{equation}}
\def \beqa{\begin{eqnarray}}
\def \eeqa{\end{eqnarray}}

\begin{document}

\title{Baryon diffusion coefficient of the strongly interacting medium}

\author{Tribhuban Parida}
\email{tribhubanp18@iiserbpr.ac.in}
\author{Sandeep Chatterjee}
\email{sandeep@iiserbpr.ac.in}

\affiliation{Department of Physical Sciences,\\
Indian Institute of Science Education and Research Berhampur,\\ 
Transit Campus (Govt ITI), Berhampur-760010, Odisha, India}

\begin{abstract}

We propose that the transverse momentum ($p_T$) differential splitting of directed flow ($\Delta v_1$) between 
proton and anti-proton can serve as a sensitive observable to extract the baryon diffusion 
coefficient ($\kappa_B$) of the hot and dense strongly interacting matter produced in relativistic heavy ion 
collisions. We use relativistic dissipative hydrodynamics framework with Glauber model based initial condition for 
the energy as well as baryon deposition that is calibrated to capture the rapidity dependence of charged particle 
multiplicity, net proton yield as well as the elusive $v_1$ splitting between proton and anti-proton. We employ the 
commonly used kinetic theory motivated ansatz: 
$\kappa_B= C_B \frac{n_B}{T} \left( \frac{1}{3} \text{coth}\left(\frac{\mu_B}{T} \right) - 
                    \frac{n_BT}{\epsilon+P}   \right)$ where $n_B$, $\epsilon$, $P$, $T$ and $\mu_B$ 
are baryon number density, energy density, pressure, temperature and baryon chemical potential respectively 
while $C_B$ is an arbitrary constant which is largely unknown for the Quantum Chromodynamics (QCD) medium.
We find that the variation of $\Delta v_1$ with $p_T$ is strongly influenced by the choice of $C_B$. Further, we 
find that the recent STAR measurement of the centrality dependence of the rapidity slope of $\Delta v_1$ 
prefers $0.5<C_B<1.5$.

\end{abstract}

\maketitle

\section{Introduction}
As a result of a relativistic heavy ion collision (RHIC), a part of the total energy-momentum, net baryon 
and electric charges carried by the colliding nuclei are deposited in the overlap region. Relativistic 
dissipative hydrodynamics has provided a framework that has been largely successful in comprehending the 
phenomenology of RHIC~\cite{Romatschke:2007mq,Schenke:2011bn,Denicol:2018wdp,Karpenko:2015xea,Shen:2012vn}. 
The three main ingredients that influence the relativistic dissipative hydrodynamics 
response are the initial conditions of the conserved quantities, the equation of state of the QCD medium and  
the medium transport coefficients that characterise the dissipative currents in the system. At the LHC energies, 
the central rapidity region is mostly free of net proton~\cite{ALICE:2013mez}. Hence it is a standard practice to consider only the 
evolution of the energy-momentum tensor ($T^{\mu\nu}$) within the relativistic dissipative hydrodynamics framework and 
neglect the other conserved charges. Shear ($\eta$) and bulk ($\zeta$) viscosities are the main transport coefficients that arise as 
a hydrodynamic response of the medium to gradients of $T^{\mu\nu}$. Reliable first principle estimates of $\eta$ and 
$\zeta$ are still missing and hence it it is a standard practice to treat the transport coefficients as unknown model parameters 
to be fixed from data to model comparisons once a sensitive observable is identified. In fact, one of the hallmark results 
of RHIC phenomenology is the extraction of $\eta$ of the QCD medium by comparing model estimates to measurement 
of the elliptic flow~\cite{Romatschke:2007mq,Karpenko:2015xea, Gotz:2022naz, Niemi:2015qia,Schenke:2011bn,Denicol:2015nhu,DerradideSouza:2015kpt,Gale:2013da,Heinz:2013th,Bernhard:2019bmu}. 
There have been also attempts to extract the bulk viscosity of the QCD medium as well~\cite{Rose:2014fba,Ryu:2015vwa,Bernhard:2019bmu}.

As one dials down the colliding energy in the Beam Energy Scan (BES) program, the net proton yield indicates 
that the matter produced in the central rapidity region becomes progressively richer in net baryon density~\cite{STAR:2017sal,BRAHMS:2003wwg,BRAHMS:2009wlg,NA49:2010lhg}. 
This suggests that unlike at higher beam energies, one can not neglect the evolution of the baryon density. 
This calls for incorporating the evolution of the baryon conserved charge within the relativistic dissipative hydrodynamics 
framework~\cite{Denicol:2018wdp,Wu:2021fjf,Cimerman:2023hjw,De:2022yxq,Monnai:2012jc}. Such a framework would 
require the QCD equation of state at non-zero baryon densities, an initial condition for the baryon density and in addition to 
$\eta$ and $\zeta$, a new transport coefficient associated with the medium response to baryon inhomogeneities of the 
medium, the baryon diffusion coefficient ($\kappa_B$). The QCD equation of state for the relevant temperature ($T$) and 
baryon chemical potential ($\mu_B$) ranges that are explored in Au+Au for $\sNN=$ 7.7 - 200 GeV are now available from 
the state of the art Lattice QCD computations~\cite{HotQCD:2014kol,HotQCD:2012fhj, Ding:2015fca,Bazavov:2017dus,Borsanyi:2013bia,Borsanyi:2018grb,Bazavov:2020bjn,Noronha-Hostler:2019ayj,Monnai:2019hkn}. 
We have recently proposed a suitable initial condition of the baryon density for the BES energies that captures the data trends in 
the $v_1$ of identified hadrons, particularly the observed $v_1$ splitting in $\prp$ and $\pbar$~\cite{Parida:2022zse,Parida:2022ppj}. 
In this work, we focus on the extraction of $\kappa_B$ by model to data comparison of suitable observables that are sensitive to 
$\kappa_B$.

The constituents formed in the aftermath of
relativistic heavy ion collision may carry multiple quantum numbers related to baryon number($B$), strangeness($S$) 
and electric charge($Q$). For example, a strange quark has $B = \frac{1}{3}$, $Q = \frac{-1}{3}$, and $S = -1$. Thus, the diffusion current of each conserved charge is not only proportional to the gradient of that specific charge's chemical potential but also includes a non-zero contribution from the gradient of the other two conserved charges. Conversely, the gradient of one conserved charge can produce the diffusion current of all three conserved charges. In general, this intermixing leads to a diffusion coefficient matrix, where the diagonal terms ($\kappa_{BB}$, $\kappa_{SS}$, $\kappa_{QQ}$) are the usual charge diffusion coefficients, and the off-diagonal terms ($\kappa_{BQ}$, $\kappa_{QS}$, $\kappa_{BS}$) quantify the coupling between different conserved charge diffusion. Recent works have calculated the full diffusion matrix for both the hadronic and partonic phases using kinetic theory~\cite{Fotakis:2021diq,Rose:2020sjv,Fotakis:2019nbq,Greif:2017byw,Das:2021bkz}. The off-diagonal elements of the diffusion matrix have been found to be of similar magnitude as the diagonal elements, indicating that coupling between the diffusion currents of conserved charges should not be neglected in simulations of low-energy heavy-ion collisions. Therefore, it is important to estimate the full diffusion matrix instead of only one component, which requires the development of a hydrodynamic framework that will evolve all three conserved charges along with the energy-momentum tensor starting from a proper initial condition with an appropriate equation of state as input. One significant progress in this direction has been done by Fotakis et. al.~\cite{Fotakis:2022usk} by deriving the second-order equations of motion for the dissipative quantities in the (10 + 4N)-moment approximation.  However, as a first step, we use the hydrodynamic framework that only considers the baryon diffusion current proportional to the gradient of the baryon chemical potential~\cite{Denicol:2018wdp,Li:2018fow}. 

\section{Model}
We use a tilted initial condition for the energy density density~\cite{Bozek:2010bi}. We adopt a suitable initial condition 
of the net baryon density $n_B$ that is an admixture of contributions from both the participant as 
well as binary collision sources~\cite{Parida:2022zse,Parida:2022ppj}. This scheme is successful in describing the observed
identified hadrons $v_1$ across beam energies, in particular the measured split in $v_1$ of $\prp$ and $\pbar$~\cite{STAR:2014clz,STAR:2017okv}. 
We further evolve the $T^{\mu\nu}$ and $n_B$ hydrodynamically using the publicly available MUSIC code~\cite{Schenke:2010nt,Schenke:2011bn,Denicol:2018wdp} 
followed by particalization using iSS code~\cite{https://doi.org/10.48550/arxiv.1409.8164,https://github.com/chunshen1987/iSS} and a hadronic 
afterburner stage using the UrQMD code~\cite{Bass:1998ca, Bleicher:1999xi}. The various model parameters are calibrated by comparing the 
model results to the data of various observables like the rapidity dependence of the identified hadron yields as well as their $v_1$. The detailed 
model scheme, fixing of the model parameters as well as their values are available in Ref.~\cite{Parida:2022zse,Parida:2022ppj}. All the 
computations in this paper are done for Au+Au at $\sNN=27$ GeV.

\begin{figure}
 \begin{center}
 \includegraphics[scale=0.65]{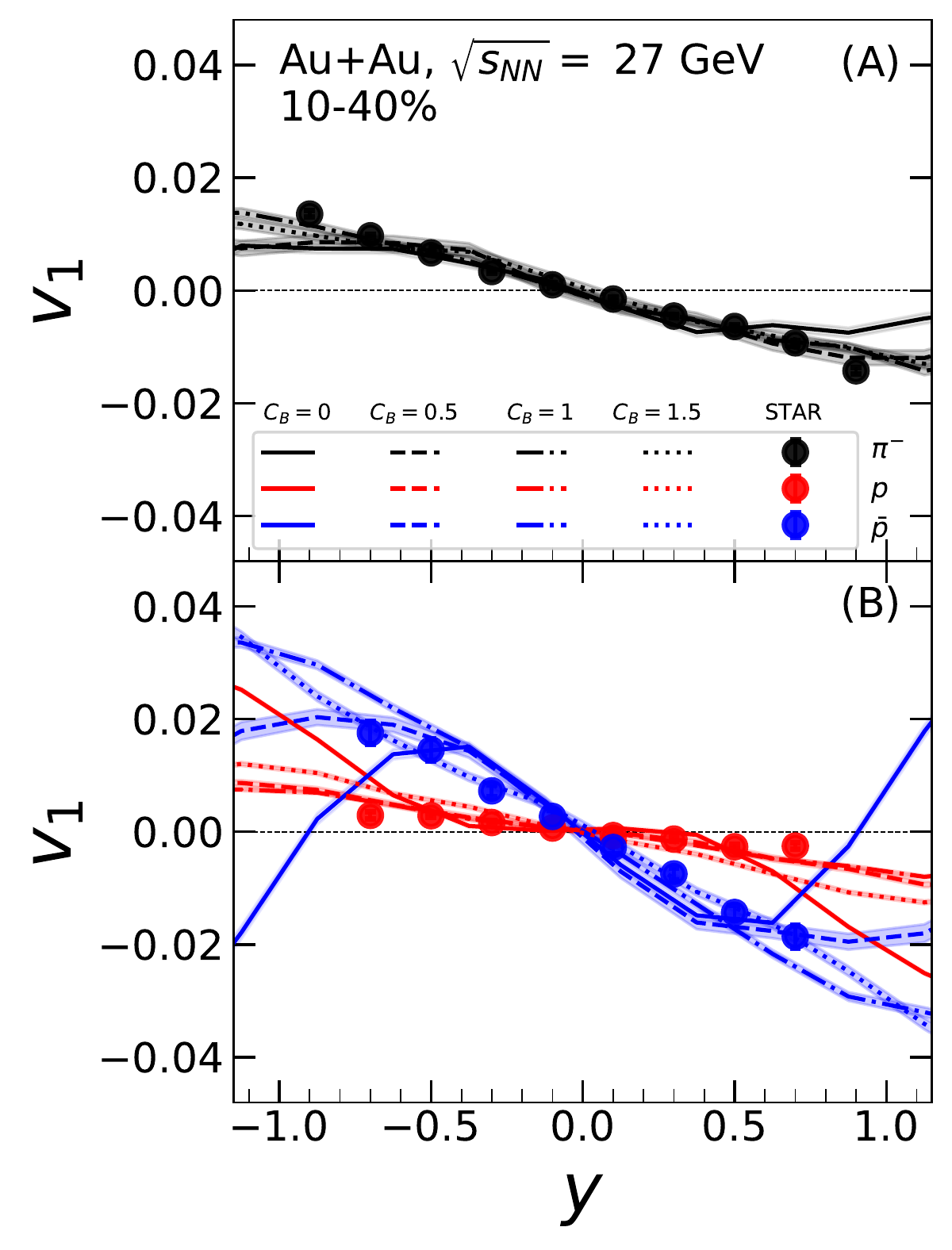}
 \caption{(Color online) $v_1$ vs $y$ is plotted for $10-40\%$ centrality Au+Au at $\sNN=27$ GeV. The effect 
 of varying strength of the baryon transport coefficient $\kappa_B$ is studied by changing $C_B$ (see 
 Eqn.~\ref{eq.kappab}). For each $C_B$ the initial condition for $\epsilon$ and $n_B$ is tuned to match the 
 STAR measurement~\cite{STAR:2014clz} of $v_1$ vs $y$ for $\pi^-$ (shown in panel A) and $\prp$ and $\pbar$ (shown in 
 panel B)}
 \label{fig.v1vsy}
 \end{center}
\end{figure}

For $\kappa_B$, we use the following ansatz based on kinetic theory approach~\cite{Denicol:2018wdp}:
\beq
\kappa_B = C_B \frac{n_B}{T} \left( \frac{1}{3} \text{coth}\left(\frac{\mu_B}{T} \right) - \frac{n_BT}{\epsilon+P}   \right)
\label{eq.kappab}
\eeq
where $n_B$, $\epsilon$, $P$, $T$ and $\mu_B$ are baryon number density, energy density, pressure, temperature and baryon chemical 
potential respectively while $C_B$ is an arbitrary constant which is largely unknown for the Quantum Chromodynamics (QCD) medium.
In this work, we demonstrate that the $p_T$ dependence of $\Delta v_1$ between proton and anti-proton provides us a sensitive probe to 
extract $\kappa_B$ of the QCD medium.

\section{RESULTS}

\begin{figure*}
 \begin{center}
 \includegraphics[scale=0.55]{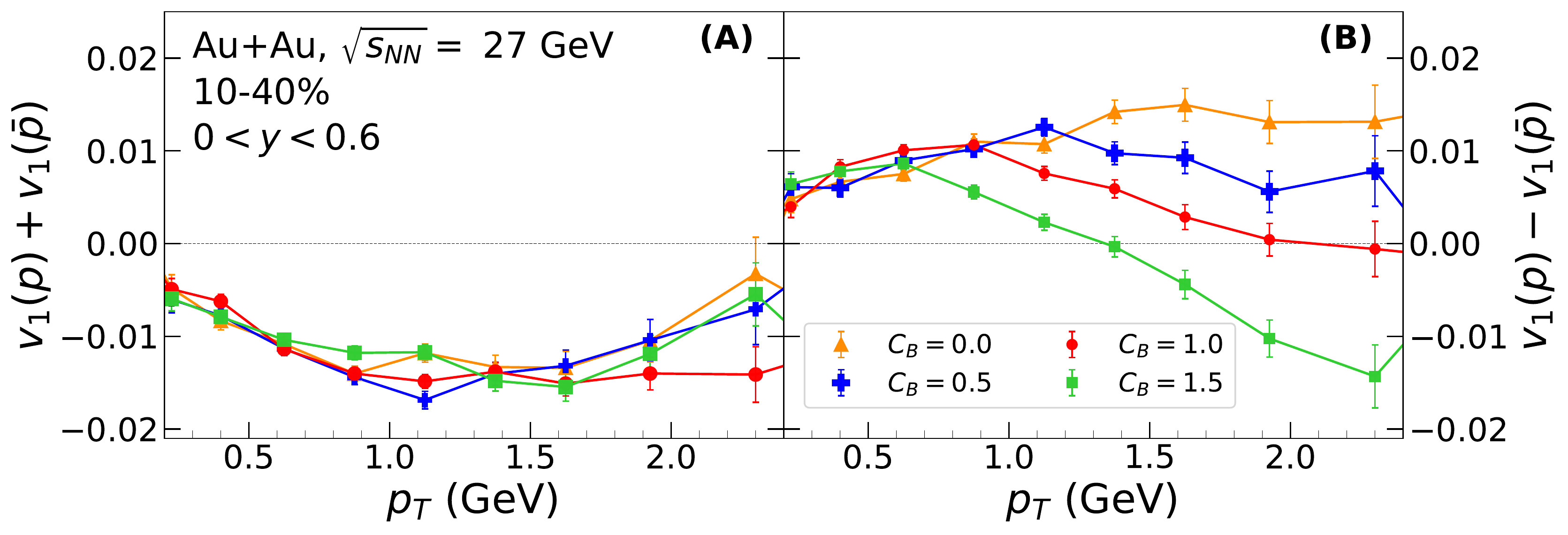}
 \caption{(Color online) The $p_T$ differential $v_1$ of $\prp$ and $\pbar$ are plotted for different choices of $C_B$ 
 that characterises the magnitude of $\kappa_B$ of the QCD medium (see Eqn.~\ref{eq.kappab}). While the sum of 
 $v_1$ of $\prp$ and $\pbar$ has a $p_T$ dependence that is insensitive to $C_B$ (shown in panel A), the $p_T$ 
 differential of their difference (shown in panel B) is sensitive to the choice of $C_B$, providing an opportunity for a 
 data driven extraction of the baryon diffusion coefficient of the QCD medium.}
 \label{fig.pT}
 \end{center}
\end{figure*}

In order to understand the phenomenological implications of $\kappa_B$, we have performed the relativistic dissipative hydrodynamic 
model simulations for different values of $\kappa_B$. We perform the calculations for four values of $C_B=$ 0, 0.5, 1 and 1.5. For every 
value of $C_B$, we choose the initial condition model parameters such that the same set of observables namely the rapidity dependence of 
the identified hadrons yield and $v_1$ match with each other as well as agree with the STAR measurement wherever available. 
In Fig.~\ref{fig.v1vsy} we have plotted the rapidity dependence of $v_1$ for $\pi^-$, $\prp$ and $\pbar$ . It is clear that 
within the current rapidity acceptance of the measurement, the rapidity dependent $v_1$ data does not discriminate between the 
different values of $C_B$ and hence cannot constrain $\kappa_B$.

Having calibrated the model parameters for all the four values of $C_B$, we now compute the $p_T$ differential $v_1$ 
of $\prp$ and $\pbar$. The rapidity acceptance used here is $0<y<0.6$. We plot the sum of the $p_T$ differential $v_1$ of 
$\prp$ and $\pbar$ in panel A of Fig.~\ref{fig.pT}. This observable seems independent of the choice of the $C_B$ value and hence 
can not be used to constrain $\kappa_B$. It will be interesting to compare this observable with data - a good agreement with data 
will be a non trivial check of the model framework independent of the value of $\kappa_B$. 

In panel B of Fig.~\ref{fig.pT}, we plot the difference of the $p_T$ differential $v_1$ of $\prp$ and $\pbar$. This observable is clearly 
very sensitive to the variation in $C_B$ and hence, $\kappa_B$. A model to data comparison of this observable can allow us to infer 
the value of $\kappa_B$ which is one of the fundamental quantities that characterise the QCD medium at non-zero baryon 
density. This is the main result of our paper.

\begin{figure}
 \begin{center}
 \includegraphics[scale=0.5]{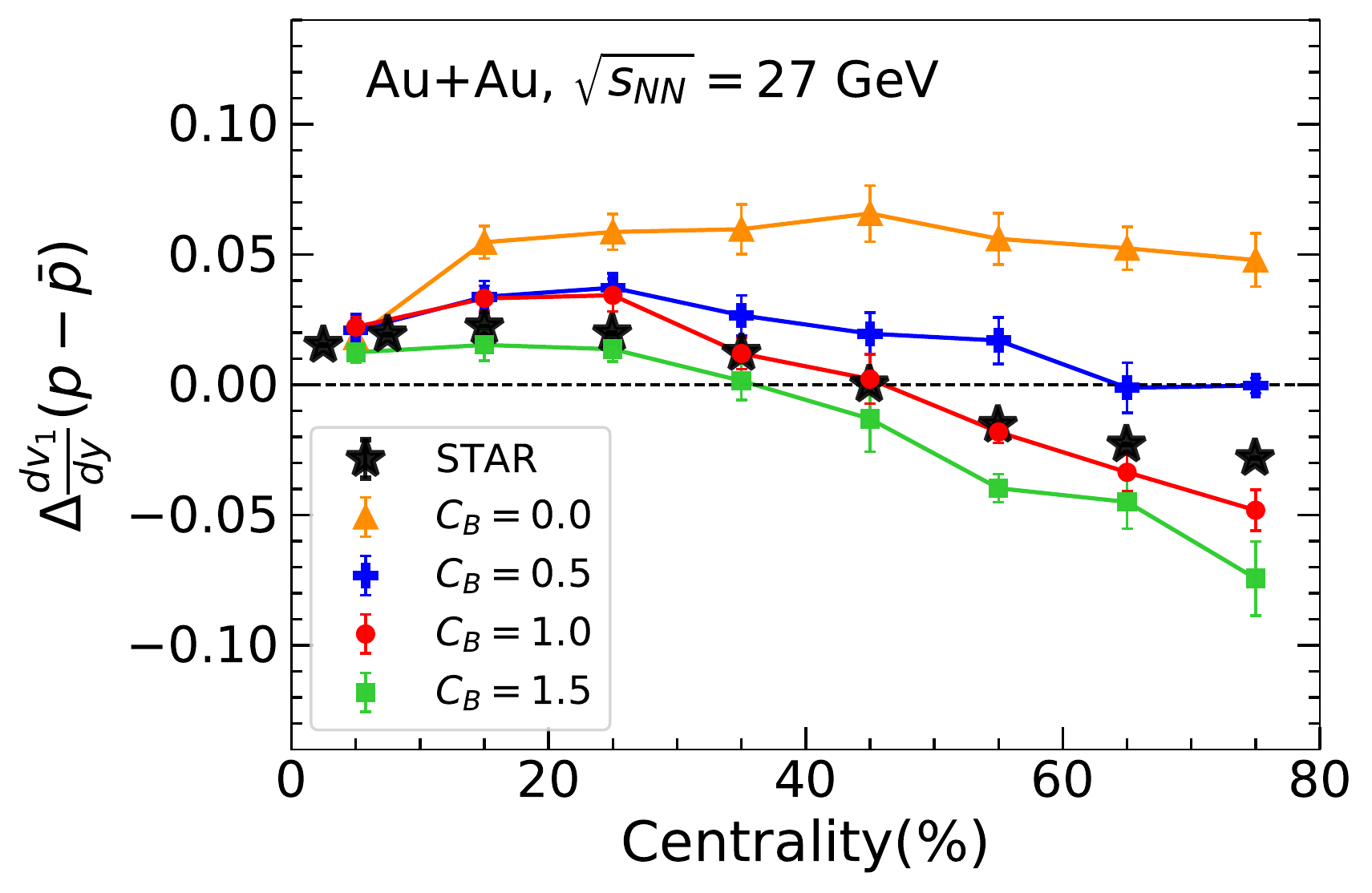}
 \caption{(Color online)The centrality dependence of the splitting in the rapidity slopes of $v_1$ of $\prp$ and 
 $\pbar$. STAR measurements~\cite{STAR:2023jdd} are compared to relativistic dissipative hydrodynamics model calculations 
 for several values of $C_B$ that controls the strength of $\kappa_B$ (see Eqn.~\ref{eq.kappab}).}
 \label{fig.cent}
 \end{center}
\end{figure}

Finally, we confront the model predictions for the different values of $C_B$ with the recent STAR measurement of the centrality 
dependence of the split in the rapidity slope 
of $v_1$ ($\Delta \frac{dv_1}{dy}$) of $\prp$ and $\pbar$ in Fig.~\ref{fig.cent}. Clearly, unlike in Fig.~\ref{fig.v1vsy} where the 
model results for all the different $C_B$ values agreed with each other, in Fig.~\ref{fig.cent} the model predictions of $\Delta \frac{dv_1}{dy}$ 
versus centralitiy vary over a wide range. This demonstrates the discriminatory power of this observable as well to extract $C_B$ and 
in turn $\kappa_B$. The model predictions for the different $C_B$ start differing from each other as one goes from central to 
peripheral collisions. The $C_B=0$ and 0.5 model predictions are always positive while those for $C_B=1$ and 1.5 undergo a 
change of sign for mid-central collisions with the larger $C_B$ results undergoing a sign change at more central collisions.
Restricting to centralities of $(0 -  40)\%$ where the possible contribution from the strong electromagnetic field to this observable 
is expected to be small\cite{Gursoy:2018yai,STAR:2023jdd} and our model results based on hydrodynamic framework are expected 
to be more reliable, we note that the data prefers $0.5 < C_B < 1.5$. This clearly underlines the significance of baryon dissipation in 
the BES phenomenology providing unique insights into the nature of the baryon rich QCD medium.

\section{SUMMARY}

We propose the measurement of $p_T$ differential split in $v_1$ of $\prp$ and $\pbar$ as a sensitive observable to probe the 
value of the baryon diffusion coefficient of the QCD medium. Further, the recent measurement by the STAR collaboration of 
the splitting in the rapidity slopes of the $v_1$ of $\prp$ and $\pbar$ versus centrality allows us to make a first estimate of 
the same: $0.5 < C_B < 1.5$ when the baryon diffusion coefficient is parametrised as in Eq.~\ref{eq.kappab}. This naturally paves 
the way towards investigations and eventual extraction of the complete $3\times3$ diffusion matrix of the QCD medium riding on 
similar observables constructed out of the relevant hadrons with appropriate quantum numbers. Our work opens 
up a novel portal to probe the QCD medium by enabling theory to data comparison of the dynamics of the QCD conserved charges.

\section{ACKNOWLEDGEMENTS}
SC acknowledges IISER Berhampur for Seed Grant.

\bibliographystyle{apsrev4-1}
\bibliography{BaryonDiffusion}

\begin{thebibliography}{51}%
\makeatletter
\providecommand \@ifxundefined [1]{%
 \@ifx{#1\undefined}
}%
\providecommand \@ifnum [1]{%
 \ifnum #1\expandafter \@firstoftwo
 \else \expandafter \@secondoftwo
 \fi
}%
\providecommand \@ifx [1]{%
 \ifx #1\expandafter \@firstoftwo
 \else \expandafter \@secondoftwo
 \fi
}%
\providecommand \natexlab [1]{#1}%
\providecommand \enquote  [1]{``#1''}%
\providecommand \bibnamefont  [1]{#1}%
\providecommand \bibfnamefont [1]{#1}%
\providecommand \citenamefont [1]{#1}%
\providecommand \href@noop [0]{\@secondoftwo}%
\providecommand \href [0]{\begingroup \@sanitize@url \@href}%
\providecommand \@href[1]{\@@startlink{#1}\@@href}%
\providecommand \@@href[1]{\endgroup#1\@@endlink}%
\providecommand \@sanitize@url [0]{\catcode `\\12\catcode `\$12\catcode
  `\&12\catcode `\#12\catcode `\^12\catcode `\_12\catcode `\%12\relax}%
\providecommand \@@startlink[1]{}%
\providecommand \@@endlink[0]{}%
\providecommand \url  [0]{\begingroup\@sanitize@url \@url }%
\providecommand \@url [1]{\endgroup\@href {#1}{\urlprefix }}%
\providecommand \urlprefix  [0]{URL }%
\providecommand \Eprint [0]{\href }%
\providecommand \doibase [0]{http://dx.doi.org/}%
\providecommand \selectlanguage [0]{\@gobble}%
\providecommand \bibinfo  [0]{\@secondoftwo}%
\providecommand \bibfield  [0]{\@secondoftwo}%
\providecommand \translation [1]{[#1]}%
\providecommand \BibitemOpen [0]{}%
\providecommand \bibitemStop [0]{}%
\providecommand \bibitemNoStop [0]{.\EOS\space}%
\providecommand \EOS [0]{\spacefactor3000\relax}%
\providecommand \BibitemShut  [1]{\csname bibitem#1\endcsname}%
\let\auto@bib@innerbib\@empty
\bibitem [{\citenamefont {Romatschke}\ and\ \citenamefont
  {Romatschke}(2007)}]{Romatschke:2007mq}%
  \BibitemOpen
  \bibfield  {author} {\bibinfo {author} {\bibfnamefont {P.}~\bibnamefont
  {Romatschke}}\ and\ \bibinfo {author} {\bibfnamefont {U.}~\bibnamefont
  {Romatschke}},\ }\href {\doibase 10.1103/PhysRevLett.99.172301} {\bibfield
  {journal} {\bibinfo  {journal} {Phys. Rev. Lett.}\ }\textbf {\bibinfo
  {volume} {99}},\ \bibinfo {pages} {172301} (\bibinfo {year} {2007})},\
  \Eprint {http://arxiv.org/abs/0706.1522} {arXiv:0706.1522 [nucl-th]}
  \BibitemShut {NoStop}%
\bibitem [{\citenamefont {Schenke}\ \emph {et~al.}(2012)\citenamefont
  {Schenke}, \citenamefont {Jeon},\ and\ \citenamefont
  {Gale}}]{Schenke:2011bn}%
  \BibitemOpen
  \bibfield  {author} {\bibinfo {author} {\bibfnamefont {B.}~\bibnamefont
  {Schenke}}, \bibinfo {author} {\bibfnamefont {S.}~\bibnamefont {Jeon}}, \
  and\ \bibinfo {author} {\bibfnamefont {C.}~\bibnamefont {Gale}},\ }\href
  {\doibase 10.1103/PhysRevC.85.024901} {\bibfield  {journal} {\bibinfo
  {journal} {Phys. Rev. C}\ }\textbf {\bibinfo {volume} {85}},\ \bibinfo
  {pages} {024901} (\bibinfo {year} {2012})},\ \Eprint
  {http://arxiv.org/abs/1109.6289} {arXiv:1109.6289 [hep-ph]} \BibitemShut
  {NoStop}%
\bibitem [{\citenamefont {Denicol}\ \emph {et~al.}(2018)\citenamefont
  {Denicol}, \citenamefont {Gale}, \citenamefont {Jeon}, \citenamefont
  {Monnai}, \citenamefont {Schenke},\ and\ \citenamefont
  {Shen}}]{Denicol:2018wdp}%
  \BibitemOpen
  \bibfield  {author} {\bibinfo {author} {\bibfnamefont {G.~S.}\ \bibnamefont
  {Denicol}}, \bibinfo {author} {\bibfnamefont {C.}~\bibnamefont {Gale}},
  \bibinfo {author} {\bibfnamefont {S.}~\bibnamefont {Jeon}}, \bibinfo {author}
  {\bibfnamefont {A.}~\bibnamefont {Monnai}}, \bibinfo {author} {\bibfnamefont
  {B.}~\bibnamefont {Schenke}}, \ and\ \bibinfo {author} {\bibfnamefont
  {C.}~\bibnamefont {Shen}},\ }\href {\doibase 10.1103/PhysRevC.98.034916}
  {\bibfield  {journal} {\bibinfo  {journal} {Phys. Rev. C}\ }\textbf {\bibinfo
  {volume} {98}},\ \bibinfo {pages} {034916} (\bibinfo {year} {2018})},\
  \Eprint {http://arxiv.org/abs/1804.10557} {arXiv:1804.10557 [nucl-th]}
  \BibitemShut {NoStop}%
\bibitem [{\citenamefont {Karpenko}\ \emph {et~al.}(2015)\citenamefont
  {Karpenko}, \citenamefont {Huovinen}, \citenamefont {Petersen},\ and\
  \citenamefont {Bleicher}}]{Karpenko:2015xea}%
  \BibitemOpen
  \bibfield  {author} {\bibinfo {author} {\bibfnamefont {I.~A.}\ \bibnamefont
  {Karpenko}}, \bibinfo {author} {\bibfnamefont {P.}~\bibnamefont {Huovinen}},
  \bibinfo {author} {\bibfnamefont {H.}~\bibnamefont {Petersen}}, \ and\
  \bibinfo {author} {\bibfnamefont {M.}~\bibnamefont {Bleicher}},\ }\href
  {\doibase 10.1103/PhysRevC.91.064901} {\bibfield  {journal} {\bibinfo
  {journal} {Phys. Rev. C}\ }\textbf {\bibinfo {volume} {91}},\ \bibinfo
  {pages} {064901} (\bibinfo {year} {2015})},\ \Eprint
  {http://arxiv.org/abs/1502.01978} {arXiv:1502.01978 [nucl-th]} \BibitemShut
  {NoStop}%
\bibitem [{\citenamefont {Shen}\ and\ \citenamefont
  {Heinz}(2012)}]{Shen:2012vn}%
  \BibitemOpen
  \bibfield  {author} {\bibinfo {author} {\bibfnamefont {C.}~\bibnamefont
  {Shen}}\ and\ \bibinfo {author} {\bibfnamefont {U.}~\bibnamefont {Heinz}},\
  }\href {\doibase 10.1103/PhysRevC.85.054902} {\bibfield  {journal} {\bibinfo
  {journal} {Phys. Rev. C}\ }\textbf {\bibinfo {volume} {85}},\ \bibinfo
  {pages} {054902} (\bibinfo {year} {2012})},\ \bibinfo {note} {[Erratum:
  Phys.Rev.C 86, 049903 (2012)]},\ \Eprint {http://arxiv.org/abs/1202.6620}
  {arXiv:1202.6620 [nucl-th]} \BibitemShut {NoStop}%
\bibitem [{\citenamefont {Abelev}\ \emph {et~al.}(2013)\citenamefont {Abelev}
  \emph {et~al.}}]{ALICE:2013mez}%
  \BibitemOpen
  \bibfield  {author} {\bibinfo {author} {\bibfnamefont {B.}~\bibnamefont
  {Abelev}} \emph {et~al.} (\bibinfo {collaboration} {ALICE}),\ }\href
  {\doibase 10.1103/PhysRevC.88.044910} {\bibfield  {journal} {\bibinfo
  {journal} {Phys. Rev. C}\ }\textbf {\bibinfo {volume} {88}},\ \bibinfo
  {pages} {044910} (\bibinfo {year} {2013})},\ \Eprint
  {http://arxiv.org/abs/1303.0737} {arXiv:1303.0737 [hep-ex]} \BibitemShut
  {NoStop}%
\bibitem [{\citenamefont {G\"otz}\ and\ \citenamefont
  {Elfner}(2022)}]{Gotz:2022naz}%
  \BibitemOpen
  \bibfield  {author} {\bibinfo {author} {\bibfnamefont {N.}~\bibnamefont
  {G\"otz}}\ and\ \bibinfo {author} {\bibfnamefont {H.}~\bibnamefont
  {Elfner}},\ }\href {\doibase 10.1103/PhysRevC.106.054904} {\bibfield
  {journal} {\bibinfo  {journal} {Phys. Rev. C}\ }\textbf {\bibinfo {volume}
  {106}},\ \bibinfo {pages} {054904} (\bibinfo {year} {2022})},\ \Eprint
  {http://arxiv.org/abs/2207.05778} {arXiv:2207.05778 [hep-ph]} \BibitemShut
  {NoStop}%
\bibitem [{\citenamefont {Niemi}\ \emph {et~al.}(2016)\citenamefont {Niemi},
  \citenamefont {Eskola},\ and\ \citenamefont {Paatelainen}}]{Niemi:2015qia}%
  \BibitemOpen
  \bibfield  {author} {\bibinfo {author} {\bibfnamefont {H.}~\bibnamefont
  {Niemi}}, \bibinfo {author} {\bibfnamefont {K.~J.}\ \bibnamefont {Eskola}}, \
  and\ \bibinfo {author} {\bibfnamefont {R.}~\bibnamefont {Paatelainen}},\
  }\href {\doibase 10.1103/PhysRevC.93.024907} {\bibfield  {journal} {\bibinfo
  {journal} {Phys. Rev. C}\ }\textbf {\bibinfo {volume} {93}},\ \bibinfo
  {pages} {024907} (\bibinfo {year} {2016})},\ \Eprint
  {http://arxiv.org/abs/1505.02677} {arXiv:1505.02677 [hep-ph]} \BibitemShut
  {NoStop}%
\bibitem [{\citenamefont {Denicol}\ \emph {et~al.}(2016)\citenamefont
  {Denicol}, \citenamefont {Monnai},\ and\ \citenamefont
  {Schenke}}]{Denicol:2015nhu}%
  \BibitemOpen
  \bibfield  {author} {\bibinfo {author} {\bibfnamefont {G.}~\bibnamefont
  {Denicol}}, \bibinfo {author} {\bibfnamefont {A.}~\bibnamefont {Monnai}}, \
  and\ \bibinfo {author} {\bibfnamefont {B.}~\bibnamefont {Schenke}},\ }\href
  {\doibase 10.1103/PhysRevLett.116.212301} {\bibfield  {journal} {\bibinfo
  {journal} {Phys. Rev. Lett.}\ }\textbf {\bibinfo {volume} {116}},\ \bibinfo
  {pages} {212301} (\bibinfo {year} {2016})},\ \Eprint
  {http://arxiv.org/abs/1512.01538} {arXiv:1512.01538 [nucl-th]} \BibitemShut
  {NoStop}%
\bibitem [{\citenamefont {Derradi~de Souza}\ \emph {et~al.}(2016)\citenamefont
  {Derradi~de Souza}, \citenamefont {Koide},\ and\ \citenamefont
  {Kodama}}]{DerradideSouza:2015kpt}%
  \BibitemOpen
  \bibfield  {author} {\bibinfo {author} {\bibfnamefont {R.}~\bibnamefont
  {Derradi~de Souza}}, \bibinfo {author} {\bibfnamefont {T.}~\bibnamefont
  {Koide}}, \ and\ \bibinfo {author} {\bibfnamefont {T.}~\bibnamefont
  {Kodama}},\ }\href {\doibase 10.1016/j.ppnp.2015.09.002} {\bibfield
  {journal} {\bibinfo  {journal} {Prog. Part. Nucl. Phys.}\ }\textbf {\bibinfo
  {volume} {86}},\ \bibinfo {pages} {35} (\bibinfo {year} {2016})},\ \Eprint
  {http://arxiv.org/abs/1506.03863} {arXiv:1506.03863 [nucl-th]} \BibitemShut
  {NoStop}%
\bibitem [{\citenamefont {Gale}\ \emph {et~al.}(2013)\citenamefont {Gale},
  \citenamefont {Jeon},\ and\ \citenamefont {Schenke}}]{Gale:2013da}%
  \BibitemOpen
  \bibfield  {author} {\bibinfo {author} {\bibfnamefont {C.}~\bibnamefont
  {Gale}}, \bibinfo {author} {\bibfnamefont {S.}~\bibnamefont {Jeon}}, \ and\
  \bibinfo {author} {\bibfnamefont {B.}~\bibnamefont {Schenke}},\ }\href
  {\doibase 10.1142/S0217751X13400113} {\bibfield  {journal} {\bibinfo
  {journal} {Int. J. Mod. Phys. A}\ }\textbf {\bibinfo {volume} {28}},\
  \bibinfo {pages} {1340011} (\bibinfo {year} {2013})},\ \Eprint
  {http://arxiv.org/abs/1301.5893} {arXiv:1301.5893 [nucl-th]} \BibitemShut
  {NoStop}%
\bibitem [{\citenamefont {Heinz}\ and\ \citenamefont
  {Snellings}(2013)}]{Heinz:2013th}%
  \BibitemOpen
  \bibfield  {author} {\bibinfo {author} {\bibfnamefont {U.}~\bibnamefont
  {Heinz}}\ and\ \bibinfo {author} {\bibfnamefont {R.}~\bibnamefont
  {Snellings}},\ }\href {\doibase 10.1146/annurev-nucl-102212-170540}
  {\bibfield  {journal} {\bibinfo  {journal} {Ann. Rev. Nucl. Part. Sci.}\
  }\textbf {\bibinfo {volume} {63}},\ \bibinfo {pages} {123} (\bibinfo {year}
  {2013})},\ \Eprint {http://arxiv.org/abs/1301.2826} {arXiv:1301.2826
  [nucl-th]} \BibitemShut {NoStop}%
\bibitem [{\citenamefont {Bernhard}\ \emph {et~al.}(2019)\citenamefont
  {Bernhard}, \citenamefont {Moreland},\ and\ \citenamefont
  {Bass}}]{Bernhard:2019bmu}%
  \BibitemOpen
  \bibfield  {author} {\bibinfo {author} {\bibfnamefont {J.~E.}\ \bibnamefont
  {Bernhard}}, \bibinfo {author} {\bibfnamefont {J.~S.}\ \bibnamefont
  {Moreland}}, \ and\ \bibinfo {author} {\bibfnamefont {S.~A.}\ \bibnamefont
  {Bass}},\ }\href {\doibase 10.1038/s41567-019-0611-8} {\bibfield  {journal}
  {\bibinfo  {journal} {Nature Phys.}\ }\textbf {\bibinfo {volume} {15}},\
  \bibinfo {pages} {1113} (\bibinfo {year} {2019})}\BibitemShut {NoStop}%
\bibitem [{\citenamefont {Rose}\ \emph {et~al.}(2014)\citenamefont {Rose},
  \citenamefont {Paquet}, \citenamefont {Denicol}, \citenamefont {Luzum},
  \citenamefont {Schenke}, \citenamefont {Jeon},\ and\ \citenamefont
  {Gale}}]{Rose:2014fba}%
  \BibitemOpen
  \bibfield  {author} {\bibinfo {author} {\bibfnamefont {J.-B.}\ \bibnamefont
  {Rose}}, \bibinfo {author} {\bibfnamefont {J.-F.}\ \bibnamefont {Paquet}},
  \bibinfo {author} {\bibfnamefont {G.~S.}\ \bibnamefont {Denicol}}, \bibinfo
  {author} {\bibfnamefont {M.}~\bibnamefont {Luzum}}, \bibinfo {author}
  {\bibfnamefont {B.}~\bibnamefont {Schenke}}, \bibinfo {author} {\bibfnamefont
  {S.}~\bibnamefont {Jeon}}, \ and\ \bibinfo {author} {\bibfnamefont
  {C.}~\bibnamefont {Gale}},\ }\href {\doibase 10.1016/j.nuclphysa.2014.09.044}
  {\bibfield  {journal} {\bibinfo  {journal} {Nucl. Phys. A}\ }\textbf
  {\bibinfo {volume} {931}},\ \bibinfo {pages} {926} (\bibinfo {year}
  {2014})},\ \Eprint {http://arxiv.org/abs/1408.0024} {arXiv:1408.0024
  [nucl-th]} \BibitemShut {NoStop}%
\bibitem [{\citenamefont {Ryu}\ \emph {et~al.}(2015)\citenamefont {Ryu},
  \citenamefont {Paquet}, \citenamefont {Shen}, \citenamefont {Denicol},
  \citenamefont {Schenke}, \citenamefont {Jeon},\ and\ \citenamefont
  {Gale}}]{Ryu:2015vwa}%
  \BibitemOpen
  \bibfield  {author} {\bibinfo {author} {\bibfnamefont {S.}~\bibnamefont
  {Ryu}}, \bibinfo {author} {\bibfnamefont {J.~F.}\ \bibnamefont {Paquet}},
  \bibinfo {author} {\bibfnamefont {C.}~\bibnamefont {Shen}}, \bibinfo {author}
  {\bibfnamefont {G.~S.}\ \bibnamefont {Denicol}}, \bibinfo {author}
  {\bibfnamefont {B.}~\bibnamefont {Schenke}}, \bibinfo {author} {\bibfnamefont
  {S.}~\bibnamefont {Jeon}}, \ and\ \bibinfo {author} {\bibfnamefont
  {C.}~\bibnamefont {Gale}},\ }\href {\doibase 10.1103/PhysRevLett.115.132301}
  {\bibfield  {journal} {\bibinfo  {journal} {Phys. Rev. Lett.}\ }\textbf
  {\bibinfo {volume} {115}},\ \bibinfo {pages} {132301} (\bibinfo {year}
  {2015})},\ \Eprint {http://arxiv.org/abs/1502.01675} {arXiv:1502.01675
  [nucl-th]} \BibitemShut {NoStop}%
\bibitem [{\citenamefont {Adamczyk}\ \emph {et~al.}(2017)\citenamefont
  {Adamczyk} \emph {et~al.}}]{STAR:2017sal}%
  \BibitemOpen
  \bibfield  {author} {\bibinfo {author} {\bibfnamefont {L.}~\bibnamefont
  {Adamczyk}} \emph {et~al.} (\bibinfo {collaboration} {STAR}),\ }\href
  {\doibase 10.1103/PhysRevC.96.044904} {\bibfield  {journal} {\bibinfo
  {journal} {Phys. Rev. C}\ }\textbf {\bibinfo {volume} {96}},\ \bibinfo
  {pages} {044904} (\bibinfo {year} {2017})},\ \Eprint
  {http://arxiv.org/abs/1701.07065} {arXiv:1701.07065 [nucl-ex]} \BibitemShut
  {NoStop}%
\bibitem [{\citenamefont {Bearden}\ \emph {et~al.}(2004)\citenamefont {Bearden}
  \emph {et~al.}}]{BRAHMS:2003wwg}%
  \BibitemOpen
  \bibfield  {author} {\bibinfo {author} {\bibfnamefont {I.~G.}\ \bibnamefont
  {Bearden}} \emph {et~al.} (\bibinfo {collaboration} {BRAHMS}),\ }\href
  {\doibase 10.1103/PhysRevLett.93.102301} {\bibfield  {journal} {\bibinfo
  {journal} {Phys. Rev. Lett.}\ }\textbf {\bibinfo {volume} {93}},\ \bibinfo
  {pages} {102301} (\bibinfo {year} {2004})},\ \Eprint
  {http://arxiv.org/abs/nucl-ex/0312023} {arXiv:nucl-ex/0312023} \BibitemShut
  {NoStop}%
\bibitem [{\citenamefont {Arsene}\ \emph {et~al.}(2009)\citenamefont {Arsene}
  \emph {et~al.}}]{BRAHMS:2009wlg}%
  \BibitemOpen
  \bibfield  {author} {\bibinfo {author} {\bibfnamefont {I.~C.}\ \bibnamefont
  {Arsene}} \emph {et~al.} (\bibinfo {collaboration} {BRAHMS}),\ }\href
  {\doibase 10.1016/j.physletb.2009.05.049} {\bibfield  {journal} {\bibinfo
  {journal} {Phys. Lett. B}\ }\textbf {\bibinfo {volume} {677}},\ \bibinfo
  {pages} {267} (\bibinfo {year} {2009})},\ \Eprint
  {http://arxiv.org/abs/0901.0872} {arXiv:0901.0872 [nucl-ex]} \BibitemShut
  {NoStop}%
\bibitem [{\citenamefont {Anticic}\ \emph {et~al.}(2011)\citenamefont {Anticic}
  \emph {et~al.}}]{NA49:2010lhg}%
  \BibitemOpen
  \bibfield  {author} {\bibinfo {author} {\bibfnamefont {T.}~\bibnamefont
  {Anticic}} \emph {et~al.} (\bibinfo {collaboration} {NA49}),\ }\href
  {\doibase 10.1103/PhysRevC.83.014901} {\bibfield  {journal} {\bibinfo
  {journal} {Phys. Rev. C}\ }\textbf {\bibinfo {volume} {83}},\ \bibinfo
  {pages} {014901} (\bibinfo {year} {2011})},\ \Eprint
  {http://arxiv.org/abs/1009.1747} {arXiv:1009.1747 [nucl-ex]} \BibitemShut
  {NoStop}%
\bibitem [{\citenamefont {Wu}\ \emph {et~al.}(2022)\citenamefont {Wu},
  \citenamefont {Qin}, \citenamefont {Pang},\ and\ \citenamefont
  {Wang}}]{Wu:2021fjf}%
  \BibitemOpen
  \bibfield  {author} {\bibinfo {author} {\bibfnamefont {X.-Y.}\ \bibnamefont
  {Wu}}, \bibinfo {author} {\bibfnamefont {G.-Y.}\ \bibnamefont {Qin}},
  \bibinfo {author} {\bibfnamefont {L.-G.}\ \bibnamefont {Pang}}, \ and\
  \bibinfo {author} {\bibfnamefont {X.-N.}\ \bibnamefont {Wang}},\ }\href
  {\doibase 10.1103/PhysRevC.105.034909} {\bibfield  {journal} {\bibinfo
  {journal} {Phys. Rev. C}\ }\textbf {\bibinfo {volume} {105}},\ \bibinfo
  {pages} {034909} (\bibinfo {year} {2022})},\ \Eprint
  {http://arxiv.org/abs/2107.04949} {arXiv:2107.04949 [hep-ph]} \BibitemShut
  {NoStop}%
\bibitem [{\citenamefont {Cimerman}\ \emph {et~al.}(2023)\citenamefont
  {Cimerman}, \citenamefont {Karpenko}, \citenamefont {Tomasik},\ and\
  \citenamefont {Huovinen}}]{Cimerman:2023hjw}%
  \BibitemOpen
  \bibfield  {author} {\bibinfo {author} {\bibfnamefont {J.}~\bibnamefont
  {Cimerman}}, \bibinfo {author} {\bibfnamefont {I.}~\bibnamefont {Karpenko}},
  \bibinfo {author} {\bibfnamefont {B.}~\bibnamefont {Tomasik}}, \ and\
  \bibinfo {author} {\bibfnamefont {P.}~\bibnamefont {Huovinen}},\ }\href
  {\doibase 10.1103/PhysRevC.107.044902} {\bibfield  {journal} {\bibinfo
  {journal} {Phys. Rev. C}\ }\textbf {\bibinfo {volume} {107}},\ \bibinfo
  {pages} {044902} (\bibinfo {year} {2023})},\ \Eprint
  {http://arxiv.org/abs/2301.11894} {arXiv:2301.11894 [nucl-th]} \BibitemShut
  {NoStop}%
\bibitem [{\citenamefont {De}\ \emph {et~al.}(2022)\citenamefont {De},
  \citenamefont {Kapusta}, \citenamefont {Singh},\ and\ \citenamefont
  {Welle}}]{De:2022yxq}%
  \BibitemOpen
  \bibfield  {author} {\bibinfo {author} {\bibfnamefont {A.}~\bibnamefont
  {De}}, \bibinfo {author} {\bibfnamefont {J.~I.}\ \bibnamefont {Kapusta}},
  \bibinfo {author} {\bibfnamefont {M.}~\bibnamefont {Singh}}, \ and\ \bibinfo
  {author} {\bibfnamefont {T.}~\bibnamefont {Welle}},\ }\href {\doibase
  10.1103/PhysRevC.106.054906} {\bibfield  {journal} {\bibinfo  {journal}
  {Phys. Rev. C}\ }\textbf {\bibinfo {volume} {106}},\ \bibinfo {pages}
  {054906} (\bibinfo {year} {2022})},\ \Eprint
  {http://arxiv.org/abs/2206.02655} {arXiv:2206.02655 [nucl-th]} \BibitemShut
  {NoStop}%
\bibitem [{\citenamefont {Monnai}(2012)}]{Monnai:2012jc}%
  \BibitemOpen
  \bibfield  {author} {\bibinfo {author} {\bibfnamefont {A.}~\bibnamefont
  {Monnai}},\ }\href {\doibase 10.1103/PhysRevC.86.014908} {\bibfield
  {journal} {\bibinfo  {journal} {Phys. Rev. C}\ }\textbf {\bibinfo {volume}
  {86}},\ \bibinfo {pages} {014908} (\bibinfo {year} {2012})},\ \Eprint
  {http://arxiv.org/abs/1204.4713} {arXiv:1204.4713 [nucl-th]} \BibitemShut
  {NoStop}%
\bibitem [{\citenamefont {Bazavov}\ \emph {et~al.}(2014)\citenamefont {Bazavov}
  \emph {et~al.}}]{HotQCD:2014kol}%
  \BibitemOpen
  \bibfield  {author} {\bibinfo {author} {\bibfnamefont {A.}~\bibnamefont
  {Bazavov}} \emph {et~al.} (\bibinfo {collaboration} {HotQCD}),\ }\href
  {\doibase 10.1103/PhysRevD.90.094503} {\bibfield  {journal} {\bibinfo
  {journal} {Phys. Rev. D}\ }\textbf {\bibinfo {volume} {90}},\ \bibinfo
  {pages} {094503} (\bibinfo {year} {2014})},\ \Eprint
  {http://arxiv.org/abs/1407.6387} {arXiv:1407.6387 [hep-lat]} \BibitemShut
  {NoStop}%
\bibitem [{\citenamefont {Bazavov}\ \emph {et~al.}(2012)\citenamefont {Bazavov}
  \emph {et~al.}}]{HotQCD:2012fhj}%
  \BibitemOpen
  \bibfield  {author} {\bibinfo {author} {\bibfnamefont {A.}~\bibnamefont
  {Bazavov}} \emph {et~al.} (\bibinfo {collaboration} {HotQCD}),\ }\href
  {\doibase 10.1103/PhysRevD.86.034509} {\bibfield  {journal} {\bibinfo
  {journal} {Phys. Rev. D}\ }\textbf {\bibinfo {volume} {86}},\ \bibinfo
  {pages} {034509} (\bibinfo {year} {2012})},\ \Eprint
  {http://arxiv.org/abs/1203.0784} {arXiv:1203.0784 [hep-lat]} \BibitemShut
  {NoStop}%
\bibitem [{\citenamefont {Ding}\ \emph {et~al.}(2015)\citenamefont {Ding},
  \citenamefont {Mukherjee}, \citenamefont {Ohno}, \citenamefont {Petreczky},\
  and\ \citenamefont {Schadler}}]{Ding:2015fca}%
  \BibitemOpen
  \bibfield  {author} {\bibinfo {author} {\bibfnamefont {H.~T.}\ \bibnamefont
  {Ding}}, \bibinfo {author} {\bibfnamefont {S.}~\bibnamefont {Mukherjee}},
  \bibinfo {author} {\bibfnamefont {H.}~\bibnamefont {Ohno}}, \bibinfo {author}
  {\bibfnamefont {P.}~\bibnamefont {Petreczky}}, \ and\ \bibinfo {author}
  {\bibfnamefont {H.~P.}\ \bibnamefont {Schadler}},\ }\href {\doibase
  10.1103/PhysRevD.92.074043} {\bibfield  {journal} {\bibinfo  {journal} {Phys.
  Rev. D}\ }\textbf {\bibinfo {volume} {92}},\ \bibinfo {pages} {074043}
  (\bibinfo {year} {2015})},\ \Eprint {http://arxiv.org/abs/1507.06637}
  {arXiv:1507.06637 [hep-lat]} \BibitemShut {NoStop}%
\bibitem [{\citenamefont {Bazavov}\ \emph {et~al.}(2017)\citenamefont {Bazavov}
  \emph {et~al.}}]{Bazavov:2017dus}%
  \BibitemOpen
  \bibfield  {author} {\bibinfo {author} {\bibfnamefont {A.}~\bibnamefont
  {Bazavov}} \emph {et~al.},\ }\href {\doibase 10.1103/PhysRevD.95.054504}
  {\bibfield  {journal} {\bibinfo  {journal} {Phys. Rev. D}\ }\textbf {\bibinfo
  {volume} {95}},\ \bibinfo {pages} {054504} (\bibinfo {year} {2017})},\
  \Eprint {http://arxiv.org/abs/1701.04325} {arXiv:1701.04325 [hep-lat]}
  \BibitemShut {NoStop}%
\bibitem [{\citenamefont {Borsanyi}\ \emph {et~al.}(2014)\citenamefont
  {Borsanyi}, \citenamefont {Fodor}, \citenamefont {Hoelbling}, \citenamefont
  {Katz}, \citenamefont {Krieg},\ and\ \citenamefont
  {Szabo}}]{Borsanyi:2013bia}%
  \BibitemOpen
  \bibfield  {author} {\bibinfo {author} {\bibfnamefont {S.}~\bibnamefont
  {Borsanyi}}, \bibinfo {author} {\bibfnamefont {Z.}~\bibnamefont {Fodor}},
  \bibinfo {author} {\bibfnamefont {C.}~\bibnamefont {Hoelbling}}, \bibinfo
  {author} {\bibfnamefont {S.~D.}\ \bibnamefont {Katz}}, \bibinfo {author}
  {\bibfnamefont {S.}~\bibnamefont {Krieg}}, \ and\ \bibinfo {author}
  {\bibfnamefont {K.~K.}\ \bibnamefont {Szabo}},\ }\href {\doibase
  10.1016/j.physletb.2014.01.007} {\bibfield  {journal} {\bibinfo  {journal}
  {Phys. Lett. B}\ }\textbf {\bibinfo {volume} {730}},\ \bibinfo {pages} {99}
  (\bibinfo {year} {2014})},\ \Eprint {http://arxiv.org/abs/1309.5258}
  {arXiv:1309.5258 [hep-lat]} \BibitemShut {NoStop}%
\bibitem [{\citenamefont {Borsanyi}\ \emph {et~al.}(2018)\citenamefont
  {Borsanyi}, \citenamefont {Fodor}, \citenamefont {Guenther}, \citenamefont
  {Katz}, \citenamefont {Szabo}, \citenamefont {Pasztor}, \citenamefont
  {Portillo},\ and\ \citenamefont {Ratti}}]{Borsanyi:2018grb}%
  \BibitemOpen
  \bibfield  {author} {\bibinfo {author} {\bibfnamefont {S.}~\bibnamefont
  {Borsanyi}}, \bibinfo {author} {\bibfnamefont {Z.}~\bibnamefont {Fodor}},
  \bibinfo {author} {\bibfnamefont {J.~N.}\ \bibnamefont {Guenther}}, \bibinfo
  {author} {\bibfnamefont {S.~K.}\ \bibnamefont {Katz}}, \bibinfo {author}
  {\bibfnamefont {K.~K.}\ \bibnamefont {Szabo}}, \bibinfo {author}
  {\bibfnamefont {A.}~\bibnamefont {Pasztor}}, \bibinfo {author} {\bibfnamefont
  {I.}~\bibnamefont {Portillo}}, \ and\ \bibinfo {author} {\bibfnamefont
  {C.}~\bibnamefont {Ratti}},\ }\href {\doibase 10.1007/JHEP10(2018)205}
  {\bibfield  {journal} {\bibinfo  {journal} {JHEP}\ }\textbf {\bibinfo
  {volume} {10}},\ \bibinfo {pages} {205} (\bibinfo {year} {2018})},\ \Eprint
  {http://arxiv.org/abs/1805.04445} {arXiv:1805.04445 [hep-lat]} \BibitemShut
  {NoStop}%
\bibitem [{\citenamefont {Bazavov}\ \emph {et~al.}(2020)\citenamefont {Bazavov}
  \emph {et~al.}}]{Bazavov:2020bjn}%
  \BibitemOpen
  \bibfield  {author} {\bibinfo {author} {\bibfnamefont {A.}~\bibnamefont
  {Bazavov}} \emph {et~al.},\ }\href {\doibase 10.1103/PhysRevD.101.074502}
  {\bibfield  {journal} {\bibinfo  {journal} {Phys. Rev. D}\ }\textbf {\bibinfo
  {volume} {101}},\ \bibinfo {pages} {074502} (\bibinfo {year} {2020})},\
  \Eprint {http://arxiv.org/abs/2001.08530} {arXiv:2001.08530 [hep-lat]}
  \BibitemShut {NoStop}%
\bibitem [{\citenamefont {Noronha-Hostler}\ \emph {et~al.}(2019)\citenamefont
  {Noronha-Hostler}, \citenamefont {Parotto}, \citenamefont {Ratti},\ and\
  \citenamefont {Stafford}}]{Noronha-Hostler:2019ayj}%
  \BibitemOpen
  \bibfield  {author} {\bibinfo {author} {\bibfnamefont {J.}~\bibnamefont
  {Noronha-Hostler}}, \bibinfo {author} {\bibfnamefont {P.}~\bibnamefont
  {Parotto}}, \bibinfo {author} {\bibfnamefont {C.}~\bibnamefont {Ratti}}, \
  and\ \bibinfo {author} {\bibfnamefont {J.~M.}\ \bibnamefont {Stafford}},\
  }\href {\doibase 10.1103/PhysRevC.100.064910} {\bibfield  {journal} {\bibinfo
   {journal} {Phys. Rev. C}\ }\textbf {\bibinfo {volume} {100}},\ \bibinfo
  {pages} {064910} (\bibinfo {year} {2019})},\ \Eprint
  {http://arxiv.org/abs/1902.06723} {arXiv:1902.06723 [hep-ph]} \BibitemShut
  {NoStop}%
\bibitem [{\citenamefont {Monnai}\ \emph {et~al.}(2019)\citenamefont {Monnai},
  \citenamefont {Schenke},\ and\ \citenamefont {Shen}}]{Monnai:2019hkn}%
  \BibitemOpen
  \bibfield  {author} {\bibinfo {author} {\bibfnamefont {A.}~\bibnamefont
  {Monnai}}, \bibinfo {author} {\bibfnamefont {B.}~\bibnamefont {Schenke}}, \
  and\ \bibinfo {author} {\bibfnamefont {C.}~\bibnamefont {Shen}},\ }\href
  {\doibase 10.1103/PhysRevC.100.024907} {\bibfield  {journal} {\bibinfo
  {journal} {Phys. Rev. C}\ }\textbf {\bibinfo {volume} {100}},\ \bibinfo
  {pages} {024907} (\bibinfo {year} {2019})},\ \Eprint
  {http://arxiv.org/abs/1902.05095} {arXiv:1902.05095 [nucl-th]} \BibitemShut
  {NoStop}%
\bibitem [{\citenamefont {Parida}\ and\ \citenamefont
  {Chatterjee}(2022{\natexlab{a}})}]{Parida:2022zse}%
  \BibitemOpen
  \bibfield  {author} {\bibinfo {author} {\bibfnamefont {T.}~\bibnamefont
  {Parida}}\ and\ \bibinfo {author} {\bibfnamefont {S.}~\bibnamefont
  {Chatterjee}},\ }\href@noop {} {\  (\bibinfo {year} {2022}{\natexlab{a}})},\
  \Eprint {http://arxiv.org/abs/2211.15659} {arXiv:2211.15659 [nucl-th]}
  \BibitemShut {NoStop}%
\bibitem [{\citenamefont {Parida}\ and\ \citenamefont
  {Chatterjee}(2022{\natexlab{b}})}]{Parida:2022ppj}%
  \BibitemOpen
  \bibfield  {author} {\bibinfo {author} {\bibfnamefont {T.}~\bibnamefont
  {Parida}}\ and\ \bibinfo {author} {\bibfnamefont {S.}~\bibnamefont
  {Chatterjee}},\ }\href@noop {} {\  (\bibinfo {year} {2022}{\natexlab{b}})},\
  \Eprint {http://arxiv.org/abs/2211.15729} {arXiv:2211.15729 [nucl-th]}
  \BibitemShut {NoStop}%
\bibitem [{\citenamefont {Fotakis}\ \emph {et~al.}(2021)\citenamefont
  {Fotakis}, \citenamefont {Soloveva}, \citenamefont {Greiner}, \citenamefont
  {Kaczmarek},\ and\ \citenamefont {Bratkovskaya}}]{Fotakis:2021diq}%
  \BibitemOpen
  \bibfield  {author} {\bibinfo {author} {\bibfnamefont {J.~A.}\ \bibnamefont
  {Fotakis}}, \bibinfo {author} {\bibfnamefont {O.}~\bibnamefont {Soloveva}},
  \bibinfo {author} {\bibfnamefont {C.}~\bibnamefont {Greiner}}, \bibinfo
  {author} {\bibfnamefont {O.}~\bibnamefont {Kaczmarek}}, \ and\ \bibinfo
  {author} {\bibfnamefont {E.}~\bibnamefont {Bratkovskaya}},\ }\href {\doibase
  10.1103/PhysRevD.104.034014} {\bibfield  {journal} {\bibinfo  {journal}
  {Phys. Rev. D}\ }\textbf {\bibinfo {volume} {104}},\ \bibinfo {pages}
  {034014} (\bibinfo {year} {2021})},\ \Eprint
  {http://arxiv.org/abs/2102.08140} {arXiv:2102.08140 [hep-ph]} \BibitemShut
  {NoStop}%
\bibitem [{\citenamefont {Rose}\ \emph {et~al.}(2020)\citenamefont {Rose},
  \citenamefont {Greif}, \citenamefont {Hammelmann}, \citenamefont {Fotakis},
  \citenamefont {Denicol}, \citenamefont {Elfner},\ and\ \citenamefont
  {Greiner}}]{Rose:2020sjv}%
  \BibitemOpen
  \bibfield  {author} {\bibinfo {author} {\bibfnamefont {J.-B.}\ \bibnamefont
  {Rose}}, \bibinfo {author} {\bibfnamefont {M.}~\bibnamefont {Greif}},
  \bibinfo {author} {\bibfnamefont {J.}~\bibnamefont {Hammelmann}}, \bibinfo
  {author} {\bibfnamefont {J.~A.}\ \bibnamefont {Fotakis}}, \bibinfo {author}
  {\bibfnamefont {G.~S.}\ \bibnamefont {Denicol}}, \bibinfo {author}
  {\bibfnamefont {H.}~\bibnamefont {Elfner}}, \ and\ \bibinfo {author}
  {\bibfnamefont {C.}~\bibnamefont {Greiner}},\ }\href {\doibase
  10.1103/PhysRevD.101.114028} {\bibfield  {journal} {\bibinfo  {journal}
  {Phys. Rev. D}\ }\textbf {\bibinfo {volume} {101}},\ \bibinfo {pages}
  {114028} (\bibinfo {year} {2020})},\ \Eprint
  {http://arxiv.org/abs/2001.10606} {arXiv:2001.10606 [nucl-th]} \BibitemShut
  {NoStop}%
\bibitem [{\citenamefont {Fotakis}\ \emph {et~al.}(2020)\citenamefont
  {Fotakis}, \citenamefont {Greif}, \citenamefont {Greiner}, \citenamefont
  {Denicol},\ and\ \citenamefont {Niemi}}]{Fotakis:2019nbq}%
  \BibitemOpen
  \bibfield  {author} {\bibinfo {author} {\bibfnamefont {J.~A.}\ \bibnamefont
  {Fotakis}}, \bibinfo {author} {\bibfnamefont {M.}~\bibnamefont {Greif}},
  \bibinfo {author} {\bibfnamefont {C.}~\bibnamefont {Greiner}}, \bibinfo
  {author} {\bibfnamefont {G.~S.}\ \bibnamefont {Denicol}}, \ and\ \bibinfo
  {author} {\bibfnamefont {H.}~\bibnamefont {Niemi}},\ }\href {\doibase
  10.1103/PhysRevD.101.076007} {\bibfield  {journal} {\bibinfo  {journal}
  {Phys. Rev. D}\ }\textbf {\bibinfo {volume} {101}},\ \bibinfo {pages}
  {076007} (\bibinfo {year} {2020})},\ \Eprint
  {http://arxiv.org/abs/1912.09103} {arXiv:1912.09103 [hep-ph]} \BibitemShut
  {NoStop}%
\bibitem [{\citenamefont {Greif}\ \emph {et~al.}(2018)\citenamefont {Greif},
  \citenamefont {Fotakis}, \citenamefont {Denicol},\ and\ \citenamefont
  {Greiner}}]{Greif:2017byw}%
  \BibitemOpen
  \bibfield  {author} {\bibinfo {author} {\bibfnamefont {M.}~\bibnamefont
  {Greif}}, \bibinfo {author} {\bibfnamefont {J.~A.}\ \bibnamefont {Fotakis}},
  \bibinfo {author} {\bibfnamefont {G.~S.}\ \bibnamefont {Denicol}}, \ and\
  \bibinfo {author} {\bibfnamefont {C.}~\bibnamefont {Greiner}},\ }\href
  {\doibase 10.1103/PhysRevLett.120.242301} {\bibfield  {journal} {\bibinfo
  {journal} {Phys. Rev. Lett.}\ }\textbf {\bibinfo {volume} {120}},\ \bibinfo
  {pages} {242301} (\bibinfo {year} {2018})},\ \Eprint
  {http://arxiv.org/abs/1711.08680} {arXiv:1711.08680 [hep-ph]} \BibitemShut
  {NoStop}%
\bibitem [{\citenamefont {Das}\ \emph {et~al.}(2022)\citenamefont {Das},
  \citenamefont {Mishra},\ and\ \citenamefont {Mohapatra}}]{Das:2021bkz}%
  \BibitemOpen
  \bibfield  {author} {\bibinfo {author} {\bibfnamefont {A.}~\bibnamefont
  {Das}}, \bibinfo {author} {\bibfnamefont {H.}~\bibnamefont {Mishra}}, \ and\
  \bibinfo {author} {\bibfnamefont {R.~K.}\ \bibnamefont {Mohapatra}},\ }\href
  {\doibase 10.1103/PhysRevD.106.014013} {\bibfield  {journal} {\bibinfo
  {journal} {Phys. Rev. D}\ }\textbf {\bibinfo {volume} {106}},\ \bibinfo
  {pages} {014013} (\bibinfo {year} {2022})},\ \Eprint
  {http://arxiv.org/abs/2109.01543} {arXiv:2109.01543 [nucl-th]} \BibitemShut
  {NoStop}%
\bibitem [{\citenamefont {Fotakis}\ \emph {et~al.}(2022)\citenamefont
  {Fotakis}, \citenamefont {Moln\'ar}, \citenamefont {Niemi}, \citenamefont
  {Greiner},\ and\ \citenamefont {Rischke}}]{Fotakis:2022usk}%
  \BibitemOpen
  \bibfield  {author} {\bibinfo {author} {\bibfnamefont {J.~A.}\ \bibnamefont
  {Fotakis}}, \bibinfo {author} {\bibfnamefont {E.}~\bibnamefont {Moln\'ar}},
  \bibinfo {author} {\bibfnamefont {H.}~\bibnamefont {Niemi}}, \bibinfo
  {author} {\bibfnamefont {C.}~\bibnamefont {Greiner}}, \ and\ \bibinfo
  {author} {\bibfnamefont {D.~H.}\ \bibnamefont {Rischke}},\ }\href {\doibase
  10.1103/PhysRevD.106.036009} {\bibfield  {journal} {\bibinfo  {journal}
  {Phys. Rev. D}\ }\textbf {\bibinfo {volume} {106}},\ \bibinfo {pages}
  {036009} (\bibinfo {year} {2022})},\ \Eprint
  {http://arxiv.org/abs/2203.11549} {arXiv:2203.11549 [nucl-th]} \BibitemShut
  {NoStop}%
\bibitem [{\citenamefont {Li}\ and\ \citenamefont {Shen}(2018)}]{Li:2018fow}%
  \BibitemOpen
  \bibfield  {author} {\bibinfo {author} {\bibfnamefont {M.}~\bibnamefont
  {Li}}\ and\ \bibinfo {author} {\bibfnamefont {C.}~\bibnamefont {Shen}},\
  }\href {\doibase 10.1103/PhysRevC.98.064908} {\bibfield  {journal} {\bibinfo
  {journal} {Phys. Rev. C}\ }\textbf {\bibinfo {volume} {98}},\ \bibinfo
  {pages} {064908} (\bibinfo {year} {2018})},\ \Eprint
  {http://arxiv.org/abs/1809.04034} {arXiv:1809.04034 [nucl-th]} \BibitemShut
  {NoStop}%
\bibitem [{\citenamefont {Bozek}\ and\ \citenamefont
  {Wyskiel}(2010)}]{Bozek:2010bi}%
  \BibitemOpen
  \bibfield  {author} {\bibinfo {author} {\bibfnamefont {P.}~\bibnamefont
  {Bozek}}\ and\ \bibinfo {author} {\bibfnamefont {I.}~\bibnamefont
  {Wyskiel}},\ }\href {\doibase 10.1103/PhysRevC.81.054902} {\bibfield
  {journal} {\bibinfo  {journal} {Phys. Rev. C}\ }\textbf {\bibinfo {volume}
  {81}},\ \bibinfo {pages} {054902} (\bibinfo {year} {2010})},\ \Eprint
  {http://arxiv.org/abs/1002.4999} {arXiv:1002.4999 [nucl-th]} \BibitemShut
  {NoStop}%
\bibitem [{\citenamefont {Adamczyk}\ \emph {et~al.}(2014)\citenamefont
  {Adamczyk} \emph {et~al.}}]{STAR:2014clz}%
  \BibitemOpen
  \bibfield  {author} {\bibinfo {author} {\bibfnamefont {L.}~\bibnamefont
  {Adamczyk}} \emph {et~al.} (\bibinfo {collaboration} {STAR}),\ }\href
  {\doibase 10.1103/PhysRevLett.112.162301} {\bibfield  {journal} {\bibinfo
  {journal} {Phys. Rev. Lett.}\ }\textbf {\bibinfo {volume} {112}},\ \bibinfo
  {pages} {162301} (\bibinfo {year} {2014})},\ \Eprint
  {http://arxiv.org/abs/1401.3043} {arXiv:1401.3043 [nucl-ex]} \BibitemShut
  {NoStop}%
\bibitem [{\citenamefont {Adamczyk}\ \emph {et~al.}(2018)\citenamefont
  {Adamczyk} \emph {et~al.}}]{STAR:2017okv}%
  \BibitemOpen
  \bibfield  {author} {\bibinfo {author} {\bibfnamefont {L.}~\bibnamefont
  {Adamczyk}} \emph {et~al.} (\bibinfo {collaboration} {STAR}),\ }\href
  {\doibase 10.1103/PhysRevLett.120.062301} {\bibfield  {journal} {\bibinfo
  {journal} {Phys. Rev. Lett.}\ }\textbf {\bibinfo {volume} {120}},\ \bibinfo
  {pages} {062301} (\bibinfo {year} {2018})},\ \Eprint
  {http://arxiv.org/abs/1708.07132} {arXiv:1708.07132 [hep-ex]} \BibitemShut
  {NoStop}%
\bibitem [{\citenamefont {Schenke}\ \emph {et~al.}(2010)\citenamefont
  {Schenke}, \citenamefont {Jeon},\ and\ \citenamefont
  {Gale}}]{Schenke:2010nt}%
  \BibitemOpen
  \bibfield  {author} {\bibinfo {author} {\bibfnamefont {B.}~\bibnamefont
  {Schenke}}, \bibinfo {author} {\bibfnamefont {S.}~\bibnamefont {Jeon}}, \
  and\ \bibinfo {author} {\bibfnamefont {C.}~\bibnamefont {Gale}},\ }\href
  {\doibase 10.1103/PhysRevC.82.014903} {\bibfield  {journal} {\bibinfo
  {journal} {Phys. Rev. C}\ }\textbf {\bibinfo {volume} {82}},\ \bibinfo
  {pages} {014903} (\bibinfo {year} {2010})},\ \Eprint
  {http://arxiv.org/abs/1004.1408} {arXiv:1004.1408 [hep-ph]} \BibitemShut
  {NoStop}%
\bibitem [{\citenamefont {Shen}\ \emph {et~al.}(2014)\citenamefont {Shen},
  \citenamefont {Qiu}, \citenamefont {Song}, \citenamefont {Bernhard},
  \citenamefont {Bass},\ and\ \citenamefont
  {Heinz}}]{https://doi.org/10.48550/arxiv.1409.8164}%
  \BibitemOpen
  \bibfield  {author} {\bibinfo {author} {\bibfnamefont {C.}~\bibnamefont
  {Shen}}, \bibinfo {author} {\bibfnamefont {Z.}~\bibnamefont {Qiu}}, \bibinfo
  {author} {\bibfnamefont {H.}~\bibnamefont {Song}}, \bibinfo {author}
  {\bibfnamefont {J.}~\bibnamefont {Bernhard}}, \bibinfo {author}
  {\bibfnamefont {S.}~\bibnamefont {Bass}}, \ and\ \bibinfo {author}
  {\bibfnamefont {U.}~\bibnamefont {Heinz}},\ }\href {\doibase
  10.48550/ARXIV.1409.8164} {\enquote {\bibinfo {title} {The iebe-vishnu code
  package for relativistic heavy-ion collisions},}\ } (\bibinfo {year}
  {2014})\BibitemShut {NoStop}%
\bibitem [{htt()}]{https://github.com/chunshen1987/iSS}%
  \BibitemOpen
  \href {https://github.com/chunshen1987/iSS} {\enquote {\bibinfo {title} {{The
  iSS code packge can be downloaded from
  https://github.com/chunshen1987/iSS}},}\ }\BibitemShut {NoStop}%
\bibitem [{\citenamefont {Bass}\ \emph {et~al.}(1998)\citenamefont {Bass} \emph
  {et~al.}}]{Bass:1998ca}%
  \BibitemOpen
  \bibfield  {author} {\bibinfo {author} {\bibfnamefont {S.~A.}\ \bibnamefont
  {Bass}} \emph {et~al.},\ }\href {\doibase 10.1016/S0146-6410(98)00058-1}
  {\bibfield  {journal} {\bibinfo  {journal} {Prog. Part. Nucl. Phys.}\
  }\textbf {\bibinfo {volume} {41}},\ \bibinfo {pages} {255} (\bibinfo {year}
  {1998})},\ \Eprint {http://arxiv.org/abs/nucl-th/9803035}
  {arXiv:nucl-th/9803035} \BibitemShut {NoStop}%
\bibitem [{\citenamefont {Bleicher}\ \emph {et~al.}(1999)\citenamefont
  {Bleicher} \emph {et~al.}}]{Bleicher:1999xi}%
  \BibitemOpen
  \bibfield  {author} {\bibinfo {author} {\bibfnamefont {M.}~\bibnamefont
  {Bleicher}} \emph {et~al.},\ }\href {\doibase 10.1088/0954-3899/25/9/308}
  {\bibfield  {journal} {\bibinfo  {journal} {J. Phys. G}\ }\textbf {\bibinfo
  {volume} {25}},\ \bibinfo {pages} {1859} (\bibinfo {year} {1999})},\ \Eprint
  {http://arxiv.org/abs/hep-ph/9909407} {arXiv:hep-ph/9909407} \BibitemShut
  {NoStop}%
\bibitem [{\citenamefont {STAR}(2023)}]{STAR:2023jdd}%
  \BibitemOpen
  \bibfield  {author} {\bibinfo {author} {\bibnamefont {STAR}},\ }\href@noop {}
  {\  (\bibinfo {year} {2023})},\ \Eprint {http://arxiv.org/abs/2304.03430}
  {arXiv:2304.03430 [nucl-ex]} \BibitemShut {NoStop}%
\bibitem [{\citenamefont {G\"ursoy}\ \emph {et~al.}(2018)\citenamefont
  {G\"ursoy}, \citenamefont {Kharzeev}, \citenamefont {Marcus}, \citenamefont
  {Rajagopal},\ and\ \citenamefont {Shen}}]{Gursoy:2018yai}%
  \BibitemOpen
  \bibfield  {author} {\bibinfo {author} {\bibfnamefont {U.}~\bibnamefont
  {G\"ursoy}}, \bibinfo {author} {\bibfnamefont {D.}~\bibnamefont {Kharzeev}},
  \bibinfo {author} {\bibfnamefont {E.}~\bibnamefont {Marcus}}, \bibinfo
  {author} {\bibfnamefont {K.}~\bibnamefont {Rajagopal}}, \ and\ \bibinfo
  {author} {\bibfnamefont {C.}~\bibnamefont {Shen}},\ }\href {\doibase
  10.1103/PhysRevC.98.055201} {\bibfield  {journal} {\bibinfo  {journal} {Phys.
  Rev. C}\ }\textbf {\bibinfo {volume} {98}},\ \bibinfo {pages} {055201}
  (\bibinfo {year} {2018})},\ \Eprint {http://arxiv.org/abs/1806.05288}
  {arXiv:1806.05288 [hep-ph]} \BibitemShut {NoStop}%
\end{thebibliography}%

\end{document}